\documentstyle[prl,aps,twocolumn,epsfig,times]{revtex}
\def\be{\begin{equation}}
\def\ee{\end{equation}}
\def\bea{\begin{eqnarray}}
\def\eea{\end{eqnarray}}
\begin{document}

\title{Hamiltonian models of multiphoton processes and four--photon \\ 
       squeezed states via nonlinear canonical transformations}
\author{Silvio De Siena, Antonio Di Lisi, and Fabrizio Illuminati}

\address{Dipartimento di Fisica, Universit\`a di Salerno, 
         INFM--Unit\`a di Salerno, and INFN--Sezione di Napoli, \\
         Gruppo Collegato di Salerno, via S. Allende, I--84081 
         Baronissi (Salerno) Italy}

\date{\today}
\maketitle

\begin{abstract}
We introduce nonlinear canonical
transformations that yield effective 
Hamiltonians of 
multiphoton down conversion processes, and 
we define the associated non--Gaussian
multiphoton squeezed states as the
coherent states of the multiphoton 
Hamiltonians.
We study in detail the four--photon
processes and the associated
non--Gaussian four--photon
squeezed states. 
The realization of squeezing,
the behavior of the field statistics, and 
the structure of the phase space distributions 
show that these states realize a natural 
four--photon generalization
of the two--photon squeezed states.  
\end{abstract}

\vspace{0.2cm}


\vspace{0.2cm}

Squeezed states \cite{yuen} 
represent a remarkable improvement
in interferometry, providing
an example of clearly nonclassical
states whose experimental realization is more 
readily accessible compared to that, e.g.,
of the number states.
As is well known, squeezed states are minimum uncertainty
states with unbalanced quantum noise on conjugate
observables, tuned by the squeezing parameter. 
They are typically generated by two-photon
parametric down conversion processes in media with
second--order optical nonlinearity.
In these systems, at variance with the case
of one--photon processes,
the modes of the electromagnetic field are
not independently excited but instead
coupled in pairs. As a consequence, the 
two--photon process is more adequately 
described by the field quadratures rather than by
the single field mode operator (the annihilation
operator) \cite{caves}.
In recent years, different types of nonclassical
states and generalizations of the squeezed states 
have been studied intensively
in quantum optics as well as in atomic physics and in
many other physical systems \cite{perelomov}.
A challenging, yet so far elusive goal 
has been to find a multiphoton generalization 
of the effective two--photon Hamiltonian 
model for degenerate down conversion 
processes \cite{rasetti}. 
Multiphoton processes are in fact
becoming of current interest 
in the study of the foundations of quantum 
mechanics. 
In particular, four--photon processes can be implemented
by entangling pairs of two--photon squeezed states 
in a beam splitter \cite{weinfurter}, while
high fidelity teleportation and the realization
of EPR and GHZ states via four--photon processes
have been recently demonstrated 
experimentally \cite{zeilinger}.

In this letter we introduce an effective 
Hamiltonian model of (degenerate)
multiphoton down conversion
processes and 
multiphoton squeezed states.
The model is
obtained by generalizing the canonical 
transformations originally exploited in
\cite{yuen}
to define the two--photon squeezed states.
The generalized transformations achieve 
the threefold goal
of preserving the canonical
commutation relations, of 
providing multiphoton down--conversion 
Hamiltonians, and of realizing 
squeezing.
We show that these transformations exist and
are obtained by adding to the linear Bogoliubov 
transformation arbitrary but sufficiently regular
nonlinear functions of one of the two
field quadratures, i.e. the 
fundamental operators associated to two--photon 
processes \cite{caves}. 
Generalized canonical transformations 
yielding multiphoton processes
have been first introduced
in \cite{noi}, with the nonlinearity 
placed on the first field quadrature $X_1$, and
the lowest order relevant case of
four--photon processes has been discussed as well.
However, the photon number distribution and
the phase space distributions
of the associated four--photon squeezed
states differ radically from 
those of the two--photon squeezed states 
already for very small values of the strength
of the nonlinearity (i.e. the strength of the 
multiphoton contributions).

Here we define a different
class of four--photon squeezed states (FPSSs).
They are obtained by introducing
a nonlinear canonical transformation with a quadratic
nonlinearity placed on the second field quadrature
$X_2$, and by antisqueezing it. 
We show that these states, although 
obtained by a method 
similar to that exploited in \cite{noi},
are physically very
distinct from those with nonlinearity on $X_1$,
and are the proper
four--photon analog of the two--photon
squeezed states concerning the relevant
physical aspects: down conversion
Hamiltonian and squeezing, field statistics, 
and phase space distributions. 

Let us consider a single mode
$a$ of the electromagnetic field.
Canonical transformations involving higher powers
of $a$ and $a^{\dagger}$
can be defined, although not directly
in terms of the field modes, but rather
of the field quadratures.
There are only two possible one--mode generalized canonical 
transformations, and they read:
\be
b_{i} = \mu a 
+ \nu a^{\dagger} + \gamma F(X_{i}) \, , \; \; \; i = 1, 2.
\label{tnl}
\ee
Here, $a, a^{\dagger}$ denote the one--mode fundamental 
canonical variables ($[a, a^{\dagger}] = 1$), $b_{i}$ 
denotes the transformed mode, $F$ is a
sufficiently regular, Hermitian 
function of one of the field quadratures
$X_{1} = (a + a^{\dagger})/\sqrt{2}, \; \; $
$X_{2} = - \imath (a - a^{\dagger})/\sqrt{2}$,
and $\mu$, $\nu$, $\gamma$ are complex coefficients,
where $\gamma$ is the ``coupling'' parameter 
measuring the
strength of the nonlinearity.
One transformation is generated
by $F(X_{1})$, the other by $F(X_{2})$.
Accordingly, the canonical constraint
$[b_{i},b_{i}^{\dagger}]=1$
on the transformed modes yields
two different sets of conditions on the 
coefficients of the transformations. 
The first condition is common to the
two sets; it is the standard
Bogoliubov constraint $|\mu|^{2} - |\nu|^{2} = 1$. 
The second condition,
which remarkably does not depend on the
form of the operatorial function $F$, 
is either
$Re (\mu \gamma^{*} - \nu^{*} \gamma) = 0$,
or $Im (\mu \gamma^{*} - \nu^{*} \gamma) = 0$,
depending on whether the nonlinearity is placed 
on $X_1$ or on $X_2$, respectively.
Here $Re(\cdot)$, $Im(\cdot)$ denote the real and
the imaginary parts. The two transformations (\ref{tnl})
yield two different one--mode multiphoton Hamiltonians
$H_{F}^{I} = b_{1}^{\dagger}b_{1} + 1/2$ and
$H_{F}^{II} = b_{2}^{\dagger}b_{2} + 1/2$.
It is instructive to write them
in terms of the field quadratures.
Let us denote by $X_{i}$ the quadrature
argument of the nonlinear function
$F$, and by $X_{j}$ the remaining quadrature
($i,j = 1, 2 \; , \; i \neq j$).
Without loss of generality we adopt 
the parameterization
$\mu = \cosh{r} 
\,\, , \,\, \nu = \sinh{r}e^{2 \imath \phi}$,
and we choose $\phi = 0$.
We may then write:
\be
H_{F} = \frac{e^{2 r_{i}}}{2} X_{i}^2 + 
\frac{e^{- 2 r_{i}}}{2} [X_{j} + \sqrt{2} 
{\tilde \gamma}_{i} e^{r_{i}} F (X_{i})]^2,
\label{HamF}
\ee
where $r_{1} = r$, and $r_{2} = - r$. 
The coefficient ${\tilde \gamma}_{1} = Im (\gamma)$,
as in this case the canonical 
constraints imply $\gamma$ imaginary,
while ${\tilde \gamma}_{2} = \gamma$,
as in this case the canonical constraints
imply $\gamma$ real.
Finally, $H_{F} = H_{F}^{I}$ if $i=1$, and
$H_{F} = H_{F}^{II}$ if $i=2$.
It is well known that, 
for the two--photon squeezed state (TPSS),
the instance $r>0$ implies
squeezing in $X_1$ as well as
antisqueezing in $X_2$,
due to the constraint of minimum
Heisenberg uncertainty. 
In the canonical multiphoton 
extension, this is no longer true.
In fact, we see from Eq. (\ref{HamF}) that the nonlinear
function of $X_{i}$ is added to the {\it other} 
quadrature $X_{j}$.
This shows that to the same TPSS
obtained either by squeezing one quadrature or by
antisqueezing the other quadrature, there correspond
two different possible, physically distinct
multiphoton squeezed states.
Squeezing on $X_1$ is implied by
the choice $r>0$ {\it and}
nonlinearity $F$ on $X_1$,
while antisqueezing
on $X_2$ is implied still by $r>0$ {\it but}
nonlinearity $F$ moved on $X_2$.
The two distinct states obtained by these two different
nonlinear transformations reduce to the same
TPSS with squeezing
on $X_1$ (and thus
antisqueezing on $X_2$) as the parameter
$\gamma$ goes to zero.
The Hamiltonian $H_{F}$
can be written down completely in terms
of the fundamental modes $a, a^{\dagger}$
once a particular form of $F(X_{i})$ is selected.
The choice $F(X_{i}) =
(X_{i})^{n} \; , \; i = 1, 2,$ provides 
generic $2n$--photon Hamiltonians $H_{2n}$.
The 
transformation generated by $F = (X_{2})^{2}$
leads to a four--photon
Hamiltonian $H_{4}^{II}=b_{2}^{\dagger}b_{2} + 1/2$ 
that reads:
\bea
&& H_{4}^{II} = (\cosh{2r} + 3 \gamma^2) a^{\dagger}a +
{\sinh}^{2}r + \frac{3}{4} \gamma^2 + \frac{1}{2} \nonumber \\
&& + \frac{1}{2} \gamma e^r (a^{\dagger} + a) +
\frac{1}{2} (\sinh{2r} - 3 \gamma^2) (a^{\dagger 2} + a^2 )
\nonumber \\
&& +  \frac{1}{2} \gamma e^r (a^{\dagger 2}a 
+ a^{\dagger}a^2 ) -
\frac{1}{2} \gamma e^r (a^{\dagger 3} + a^3 ) \nonumber \\
&& + \frac{3}{2} \gamma^2  a^{\dagger 2}a^2 -
\gamma^2  (a^{\dagger 3}a + a^{\dagger}a^3 )
+ \frac{1}{4} \gamma^2 (a^{\dagger 4}  + a^4 ) \, .
\label{Ham4}
\eea
We see that $H_{4}^{II}$ contains explicitly 
two--, three--, and four--photon 
degenerate parametric down conversion terms.
The multiphoton squeezed states are
the eigenvectors 
$\{|\beta \rangle_{\gamma, F}\}$ of the transformed
annihilation operator $b$ (here $b$ stands generically
for either $b_{1}$ or $b_{2}$), 
with eigenvalue $\beta$, and 
are the coherent states with respect to the vacuum state 
$|0 \rangle_{\gamma, F}$ of the 
generic multiphoton Hamiltonian:
$
|\beta \rangle_{\gamma, F} = D(\beta ) |0 \rangle_{\gamma, F},
$ 
where $D(\beta ) = 
\exp{(\beta b^{\dagger}  - \beta^{*} b)}$ 
denotes the Glauber unitary displacement
operator. The vectors $\{|\beta \rangle_{\gamma, F}\}$ 
form an overcomplete set and resolve the identity:
$1/\pi \int d^{2} \beta \; 
|\beta \rangle_{\gamma, F} \, {_{F, \gamma}\langle} \beta | = I$. 
If, as for the standard squeezed states,
we choose $\beta = \mu \alpha + \nu \alpha^{*}$, 
where $\alpha = \alpha_1 +
\imath \alpha_2$ is the coherent amplitude,
and again $\phi = 0$, the
canonical transformations (\ref{tnl}) 
are implemented by the unitary operators
$U_{i} = \exp{[\imath e^{r_{i}} {\tilde \gamma}_{i} 
G(X_{i})]} D(\alpha)S(r)$. Here
$S(r) = \exp{[r(a^{\dagger 2} - a^{2})]}$
denotes the single--mode squeezing operator
for $\phi = 0$, and
$G(x) = \int_{0}^{x} F(y) dy.$
The states $|\beta \rangle_{\gamma, F}$ are not
minimum uncertainty states, but they approximate the
minimum uncertainty on the characteristic time scale ${\tilde r}^{- 1}$
provided by the time dependence $r(t) = {\tilde r} \, t$ 
of the squeezing parameter \cite{noi}.
In the
representation in which 
the quadrature argument of the
function $F$ is diagonal,
the eigenvalue equation 
for the states $|\beta \rangle_{\gamma, F}$ 
is a simple first order linear 
differential equation for the wave functions
$\Psi_{\beta}^{\gamma F} (x_{i}) 
\equiv \langle x_{i}|\beta \rangle_{\gamma F} \; , \; i = 1, 2 \; .$
Integration yields
\bea
\Psi_{\beta}^{\gamma F} (x_{i}) & = & 
(\pi \sigma_{i})^{-1/4}
\exp{\left[ - \frac{ \left(
x_{i} - x_{i}^{(0)} \right)^2}{2 \sigma_{i}}\right] }
\nonumber \\
&& \nonumber \\
& \times & \exp{\{ \imath [c_{i} x_{i} + \sqrt{2} e^{r_{i}}
{\tilde \gamma}_{i} G(x_{i})]\}} \; , 
\label{nonGaussiano}
\eea 
where 
$\sigma_{i}  = e^{- 2 r_{i}}$,
$x_{1}^{(0)} = \sqrt{2} \alpha_{1}$,
$x_{2}^{(0)} = - \sqrt{2} \alpha_{2}$, 
$c_{1} = \sqrt{2} \alpha_{2}$,
$c_{2} = - \sqrt{2} \alpha_{1}$.
The multiphoton squeezed
states (\ref{nonGaussiano}) are 
non--Gaussian because of the non--quadratic
term $G(x_{i})$ appearing in the phase of
the wave function. However, the probability 
density $|\Psi_{\beta}^{\gamma F} (x_{i})|^{2}$
is Gaussian in the
representation in which the quadrature 
argument of the nonlinear function $F$
is diagonal. The probability density displays 
squeezing in this quadrature by the
usual factor $e^{2 r_{i}}$. 
This fact further shows
the necessary link between squeezing and
nonlinearity for the multiphoton
states. 
From now on we specialize to the two possible
four--photon cases, respectively associated
to $F=(X_{1})^{2}$ and to $F=(X_{2})^{2}$.
We work explicitly in the representation
in which the first quadrature $X_1$ is
diagonal. In this representation we can
compare congruently the two different
four--photon squeezed states (FPSSs), namely
$\Psi_{4}^{I}(x)$ associated to $F=(X_{1})^{2}$,
and $\Psi_{4}^{II}(x)$ 
associated to $F=(X_{2})^{2}$:
\bea
\Psi_{4}^{I}(x) & = &
\left( \pi e^{-2r} \right)^{-1/4} \exp{\left[ -
\frac{e^{2r}}{2}\left( x - \sqrt{2}\alpha_{1}
\right)^{2}
\right] } \nonumber \\
& \times & \exp{\left\{ \imath \sqrt{2} \left[ \alpha_{2}x
+ e^{-r}Im(\gamma)\frac{x^{3}}{3}\right] \right\} } \, , 
\label{PsiI}
\eea

\be
\Psi_{4}^{II}(x) = N^{- 1/2} \exp{(kx)}
Ai \left[\frac{lx + m}{l^{2/3}}\right] \, . 
\label{Airy}
\ee
In Eq. (\ref{Airy}) $N$ is the normalization factor,
$Ai[y]$ denotes the Airy function that
goes to zero as $y \rightarrow \infty$,
$l = {e^r}/2 \gamma$; 
$m = {e^{- 2 r}}/16  \gamma^2 - {\beta }/{\gamma }$;
$k = {e^{- r}}/4 \gamma$. 
The states (\ref{PsiI}) 
and (\ref{Airy}) are both non--Gaussian.
However, while 
$|\Psi_{4}^{I}(x)|^{2}$ is Gaussian,
$|\Psi_{4}^{II}(x)|^2$ 
is not. Its behavior is shown in Fig. 1.
\begin{figure}
\begin{center}
\includegraphics*[height=4.5cm,width=8cm]{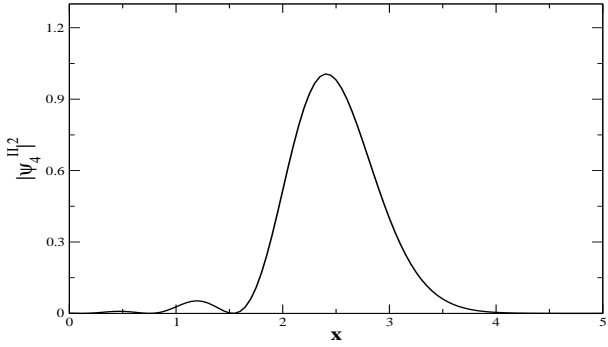}
\end{center}
\caption{The probability density $|\Psi_{4}^{II}(x)|^{2}$
of the FPSS (\ref{Airy}) 
with amplitude $\beta =3\sqrt{2}$, 
squeezing $r=0.8$, and coupling
$\gamma =0.14$.}\label{fig01}
\end{figure}

We now analyze the statistical properties
of the state (\ref{Airy}). In Fig. 2 we plot the
photon number distribution $P(n)$
of the TPSS ($\gamma=0$) 
and of the two FPSSs
$\Psi_{4}^{I}$ and $\Psi_{4}^{II}$.
\begin{figure}
\begin{center}
\includegraphics*[height=5cm,width=8cm]{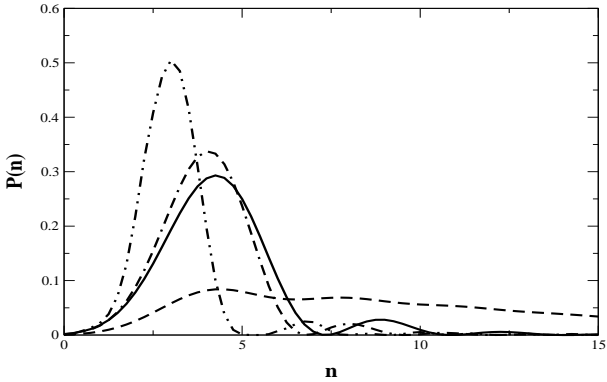}
\end{center}
\caption{The photon number distribution $P(n)$
at $\beta =3\sqrt{2}$ and $r=0.8$ for the
FPSS (\ref{Airy})
with $\gamma =0.1$ (dot--dashed line);
for the same state with $\gamma =0.5$ 
(doubledot--dashed line); for the 
FPSS (\ref{PsiI})
with $\gamma =0.05$ (dashed line); and
for the TPSS (full
line).}\label{fig02}
\end{figure}

We see from Fig. 2 that 
the form of $P(n)$ 
for the state $\Psi_{4}^{II}$
is similar to that of the TPSS
and is very stable as a function of 
the nonlinear coupling $\gamma$. The only
significant difference is that,
for larger values of the coupling the 
peak of $P(n)$ moves to the left, due to the growing
influence of $\gamma$ on the 
mean number of photons.
We see instead that $P(n)$ for the state
$\Psi_{4}^{I}$ is unstable and is 
strongly deformed already for very
small values of the nonlinear
coupling.
The profoundly different behavior
of $P(n)$ for the states $\Psi_{4}^{I}$ and
$\Psi_{4}^{II}$ is deeply rooted
in the structure of their
respective quasiprobability distributions
in phase space. 
In Fig. 3 we plot the Wigner function
for the state $\Psi_{4}^{II}$,
while in Fig. 4 we plot it
for the state $\Psi_{4}^{I}$.
The nonclassical nature of the  
state $\Psi_{4}^{II}$ is more
pronounced. Its Wigner function
is deformed and attains also negative values.
However, it is neither rotated nor translated
with respect to that of the TPSS,
at variance with the Wigner function
of the state $\Psi_{4}^{I}$ which is both
translated and rotated with respect to that 
of the TPSS.
This difference affects crucially
the properties of phase space interference. 
The latter is responsible for the
oscillations of $P(n)$ for the 
TPSS \cite{schleich}.
\begin{figure}
\begin{center}
\includegraphics*[height=6cm,width=8cm]{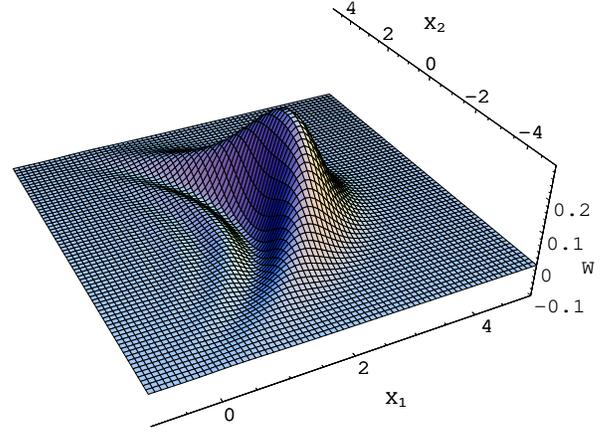}
\end{center}
\caption{The Wigner quasiprobability
distribution $W(x_{1},x_{2})$ of the FPSS (\ref{Airy}) 
with $\beta =3\sqrt{2}$, 
$r=0.8$, 
and $\gamma =0.14$.}\label{fig03}
\end{figure}

\begin{figure}
\begin{center}
\includegraphics*[height=6cm,width=8cm]{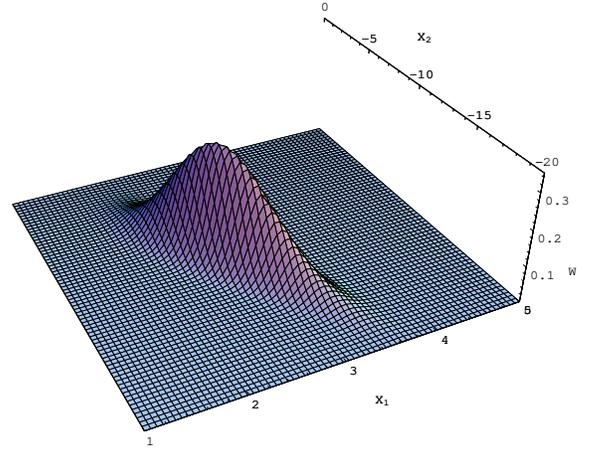}
\end{center}
\caption{The Wigner quasiprobability
distribution $W(x_{1},x_{2})$ of the FPSS (\ref{PsiI}) 
with $\beta =3\sqrt{2}$, 
$r=0.8$, 
and $\gamma =0.14$.}\label{fig04}
\end{figure}

The oscillations are 
preserved in a wide range of values of $\gamma$ 
for the state $\Psi_{4}^{II}$, exactly because its
Wigner function is neither translated nor rotated
with respect to that of the TPSS,
while they quickly disappear for the
state $\Psi_{4}^{I}$, whose Wigner function
is both translated and rotated with respect
to that of the TPSS already at small values
of $\gamma$.
In order to complete the analysis of the
statistical properties of the states
$\Psi_{4}^{I}$ and $\Psi_{4}^{II}$
we move to study their correlation functions. 
In Fig. 5 we plot the normalized 
second--order correlation
functions $g^{(2)}(0)$  
of the TPSS ($\gamma=0$) 
and of the 
states $\Psi_{4}^{I}$ and $\Psi_{4}^{II}$,
as a function of $r$.
We see that
the state $\Psi_{4}^{II}$ 
follows a behavior similar to that of
the TPSS, being exactly
Poissonian at $r=0$, slightly more 
sub--Poissonian for $r < 0.9$,
and super--Poissonian for larger values of $r$.
The value of saturation of $g^{(2)}(0)$ 
at asymptotically large values of $r$ 
is an increasing function of $\gamma$.
All these characteristics remain stable as 
$\gamma$ is varied.
At variance with this behavior, the 
state $\Psi_{4}^{I}$ 
is always super--Poissonian, even at $r=0$, and
for small values of $\gamma$.
\begin{figure}
\begin{center}
\includegraphics*[height=5cm,width=8cm]{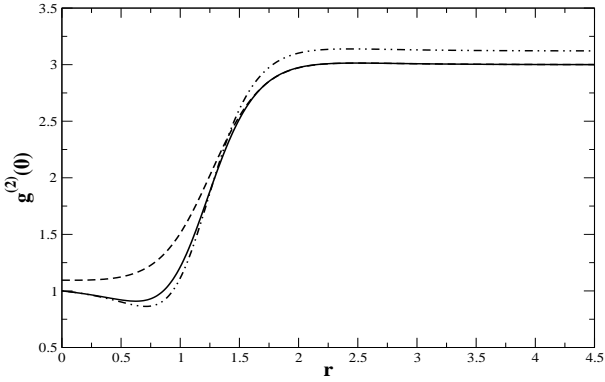}
\end{center}
\caption{The second--order
correlation $g^{(2)}(0)$
at $\beta =3\sqrt{2}$
as a function of $r$ for
the four--photon squeezed 
state (\ref{Airy}) with $\gamma =0.1$
(doubledot--dashed line); for the 
FPSS (\ref{PsiI})
with $\gamma =0.1$ (dashed line); and for
the TPSS (full line).}\label{fig05}
\end{figure}

It is of particular importance to
study the fourth--order correlation
function in order to establish
the degree of probability
of simultaneous four--photon
detection (bunching) \cite{weinfurter}.  
In Fig. 6 we plot the normalized 
fourth--order correlation
functions $g^{(4)}(0)$  
of the TPSS ($\gamma=0$) 
and of the 
states $\Psi_{4}^{I}$ and $\Psi_{4}^{II}$,
as a function of $r$.
Also in this case the FPSS $\Psi_{4}^{II}$
follows a behavior similar to that of the
TPSS. It favors four--photon
anti--bunching for values of $r < 0.8$.
For larger values of $r$ it strongly favors
four--photon bunching, and the value of saturation
of $g^{(4)}(0)$ 
grows with $\gamma$.
The state $\Psi_{4}^{I}$ instead always favors
four--photon bunching even at $r=0$ and for
small values of $\gamma$.
\begin{figure}
\begin{center}
\includegraphics*[height=5cm,width=8cm]{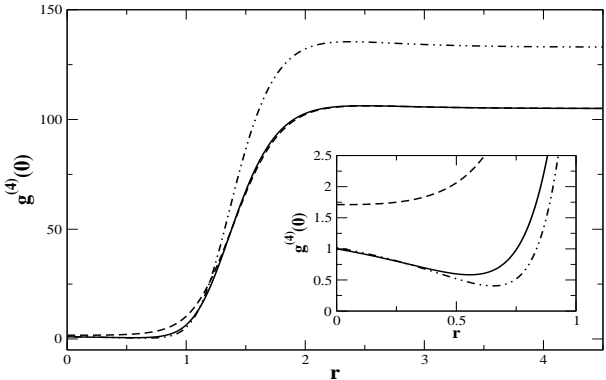}
\end{center}
\caption{The fourth--order
correlation $g^{(4)}(0)$
at $\beta =3\sqrt{2}$
as a function of $r$ for
the FPSS (\ref{Airy}) with $\gamma =0.1$
(doubledot--dashed line); for the 
FPSS (\ref{PsiI})
with $\gamma =0.1$ (dashed line); and for
the TPSS 
(full line). Inset: the same
graph in the interval $[0,1]$.}\label{fig06}
\end{figure}

In summary, we have introduced an effective
Hamiltonian model for multiphoton down conversion
processes via a nonlinear generalization of the
Bogoliubov transformation. This generalization
allows to define two different classes of
non--Gaussian
multiphoton squeezed states. We have studied
in detail the four--photon case, and we have shown 
that the four--photon squeezed
states associated to quadratic nonlinearity
in the second field quadrature and to antisqueezing
in the same quadrature are the natural four--photon
extension of the two--photon squeezed states with
respect to the Hamiltonian structure, the field
statistics, and the Wigner quasiprobability
distribution. We have considered
the instance of degenerate multiphoton
parametric down conversions. Work is in progress to 
address the description of 
nondegenerate, two-- and multimode multiphoton
processes.

\end{document}